\documentclass{iaus}
\usepackage{graphicx}

\title[X-rays from T Tauri stars] 
{Why are accreting T Tauri stars less luminous in X-rays than non-accretors?}

\author[Gregory \etal\ ]   
{S. G. Gregory$^1$%
  \thanks{Present address: SUPA, School of Physics and Astronomy, University of 
    St Andrews, St Andrews, KY16 9SS, UK.},
 K. Wood$^1$ \break \and M. Jardine$^1$}

\affiliation{$^1$SUPA, School of Physics and Astronomy, University of St Andrews,
St Andrews, KY16 9SS, UK \break email: sg64@st-andrews.ac.uk}

\pubyear{2007}
\volume{243}  
\pagerange{119--126}
\date{?? and in revised form ??}
\jname{Proceedings Title IAU Symposium}
\editors{A.C. Editor, B.D. Editor \& C.E. Editor, eds.}
\begin{document}

\maketitle

\begin{abstract}
Accreting T Tauri stars are observed to be less luminous in X-rays than non-accretors, 
an effect that has been detected in various star forming regions.  To explain this we 
have combined, for the first time, a radiative transfer code with an accretion model 
that considers magnetic fields extrapolated from surface magnetograms obtained from 
Zeeman-Doppler imaging. Such fields consist of compact magnetic regions close to the 
stellar surface, with extended field lines interacting with the disk. We study the 
propagation of coronal X-rays through the magnetosphere and demonstrate that they are 
strongly absorbed by the dense gas in accretion columns.
\keywords{radiative transfer, stars: coronae, stars: magnetic fields, 
  stars: pre--main-sequence, stars: activity, stars: formation, X-rays: stars}
\end{abstract}


\firstsection 
\section{Introduction}
Accreting T Tauri stars are observed to be less luminous in X-rays than non-accretors 
(\cite[Stelzer \& Neuh{\"a}user 2001]{ste01}; \cite[Flaccomio, Micela \& Sciortino 2003a]{fla03a}; 
\cite[Flaccomio, Damiani, Micela, \etal\ 2003c]{fla03c}; \cite[Stassun, Ardila, Barsony, \etal\ 2004]{sta04}; 
\cite[Preibisch, Kim, Favata, \etal\ 2005]{pre05}; \cite[Flaccomio, Micela \& Sciortino 2006]{fla06}; 
\cite[Telleschi, G{\"u}del, Briggs, \etal\ 2007a]{tel07a}).  Accreting stars appear to be a 
factor of $\sim 2$ less luminous, and show a larger variation 
in their X-ray activity compared to non-accreting stars (\cite[Preibisch \etal\ 2005]{pre05}). 
However, it is only 
in recent years that this result has become clear, with previous studies showing conflicting 
results (e.g. \cite{fei03} and \cite{fla03b}).  The apparent discrepancy arose from
whether stars were classified as accreting based on the detection of excess IR emission (a disk 
indicator) or the detection of accretion related emission lines.  However, with careful 
re-analysis of archival data (\cite[Flaccomio \etal\ 2003a]{fla03a}) and recent large X-ray surveys 
like the \textit{Chandra} Orion 
Ultradeep Project (COUP; \cite[Getman, Flaccomio, Broos, \etal\ 2005]{get05}) and the 
XMM-\textit{Newton} Extended Survey of the Taurus Molecular 
Cloud (XEST; \cite[G{\"u}del, Briggs, Arzner, \etal\ 2007a]{gue07a}) the result is now clear, 
namely that accreting T Tauri stars are 
observed to be, on average, less luminous in X-rays than non-accreting stars.  Although the 
difference is small it has been found consistently in various star forming regions: 
Taurus-Auriga (\cite[Stelzer \& Neuh{\"a}user 2001]{ste01}; \cite[Telleschi \etal\ 2007a]{tel07a}), 
the Orion Nebula Cluster (\cite[Flaccomio \etal\ 2003c]{fla03c}; \cite[Stassun \etal\ 2004]{sta04}; 
\cite[Preibisch \etal\ 2005]{pre05}), NGC 2264 (\cite[Flaccomio \etal\ 2003a, 2006]{fla03a,fla06}) and 
Chamaeleon I (\cite[Flaccomio \etal\ 2003a]{fla03a}).  

It should be noted, however, that such observations from CCD detectors are not very sensitive to 
X-rays that are produced in accretion shocks.  High resolution X-ray spectroscopic measurements 
have indicated emission from cool and high density plasma, most likely associated with accretion 
hot spots, in several (but not all) accreting stars (e.g. 
\cite[Telleschi, G{\"u}del, Briggs, \etal\ 2007b]{tel07b}; 
\cite[G{\"u}nther, Schmitt, Robrade, \etal\ 2007]{hans07}).  In this 
work we only consider coronal X-ray emission such as is detected by CCD measurements.   

It is not yet understood why accreting stars are under luminous in X-rays, although a few ideas have 
been put forward.  It may be related to higher extinction due to X-ray absorption by circumstellar 
disks, however the COUP results do not support this suggestion (\cite[Preibisch \etal\ 2005]{pre05}).  
It may be related to 
magnetic braking, whereby the interaction between the magnetic field of an accreting star with its 
disk slows the stellar rotation rate leading to a weaker dynamo action and therefore less X-ray 
emission; although the lack of any rotation-activity relation for T Tauri stars has ruled out this 
idea (\cite[Flaccomio \etal\ 2003c]{fla03c}; \cite[Preibisch \etal\ 2005]{pre05}; 
\cite[Briggs, G{\"u}del, Telleschi, \etal\ 2007]{bri07}).  
A third suggestion is that accretion may alter the stellar structure affecting the magnetic field 
generation process and therefore X-ray emission (\cite[Preibisch \etal\ 2005]{pre05}).  However, 
the most plausible suggestion is the 
attenuation of coronal X-rays by the dense gas in accretion columns 
(\cite[Flaccomio \etal\ 2003c]{fla03c}; \cite[Stassun \etal\ 2004]{sta04}; 
\cite[Preibisch \etal\ 2005]{pre05}; 
\cite[G{\"u}del, Telleschi, Audard, \etal\ 2007b]{gue07b}).  X-rays from the underlying corona 
may not be able to heat the material within accretion columns to a high 
enough temperature to emit in X-rays.  Field lines which have been mass-loaded with dense disk 
material may obscure the line-of-sight to the star at some rotation phases, reducing the observed 
X-ray emission.  Here we demonstrate this in a quantitative way by developing an accretion 
flow model and simulating the propagation of coronal X-rays through the stellar magnetosphere.  


\section{Realistic Magnetic Fields}
\begin{figure}
\centerline{
\scalebox{0.9}{%
\includegraphics{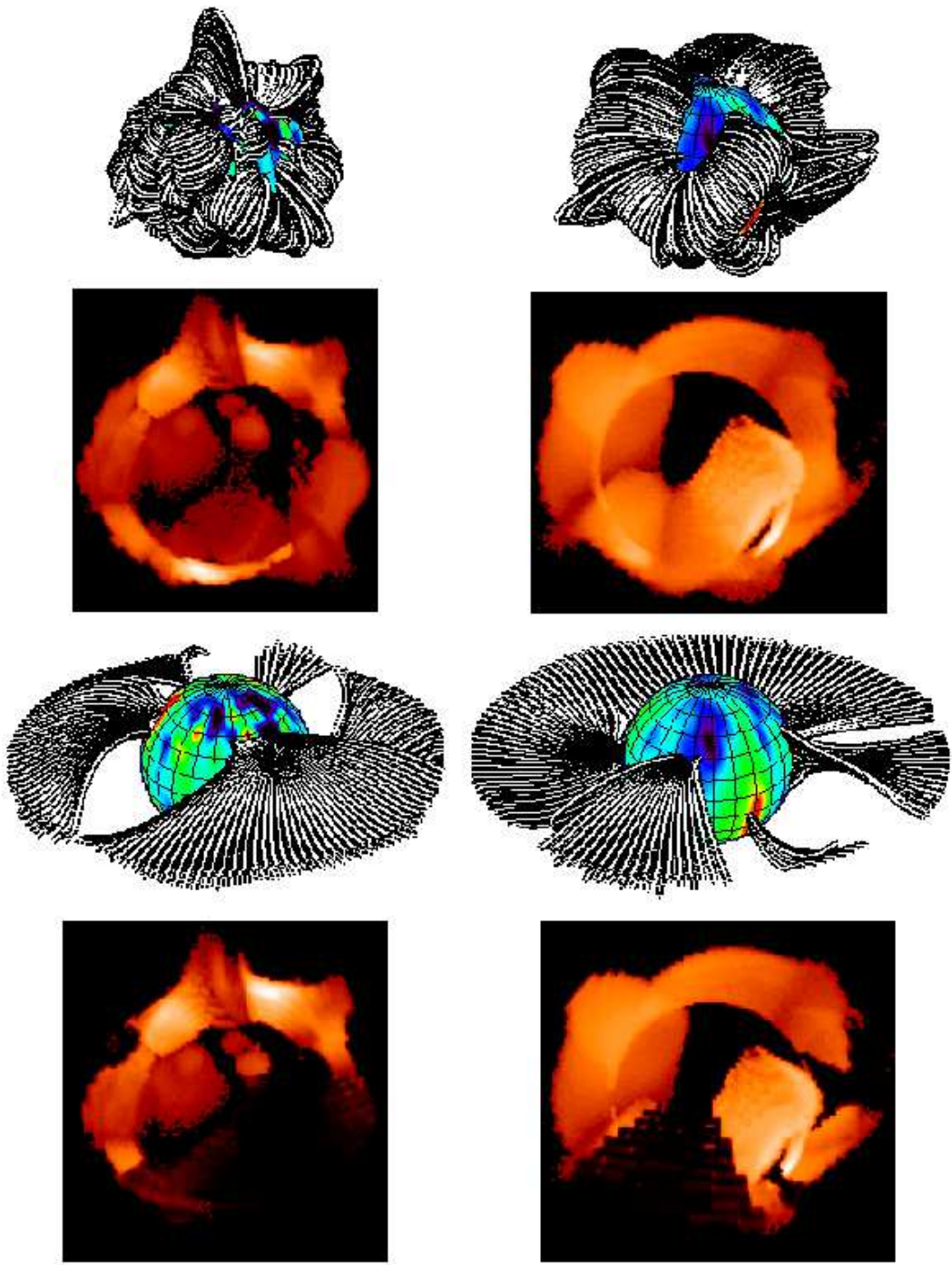}%
}
}
\caption{The model coronal (first row) T Tauri magnetic fields extrapolated from 
        the AB Dor (left-hand column) and LQ Hya (right-hand column) surface magnetograms, 
        with the corresponding X-ray corona (second row) assuming a stellar inclination of 
        $60\,^{\circ}$.  Also shown is the structure of the accreting field (third row) and the 
        X-ray emission images assuming that accretion is taking place (fourth row) - notice 
        the occulted X-ray bright regions.  For some lines-of-sight the X-ray bright regions 
        are lightly obscured by the accretion columns, reducing the observed X-ray emission.  
        For other lines-of-sight the coronal X-rays cannot penetrate the dense accreting gas.  
        The average reduction in the observed X-ray EM across an entire rotation cycle is 
        a factor of 1.4 (2.0) for the AB Dor-like (LQ Hya-like) field.  The images are not to 
        scale.  Reproduced from \cite[Gregory \etal\ (2007)]{gre07}.}
    \label{coronae}
\end{figure}

In order to model the coronae of T Tauri stars we need to assume something about the form of the 
magnetic field.  Observations suggest it is compact and inhomogeneous and may vary not only with 
time on each star, but also from one star to the next. To capture this behaviour, we use as 
examples the field structures of two different main sequence stars, LQ Hya and AB Dor determined 
from Zeeman-Doppler imaging (\cite[Donati, Cameron, Semel, \etal\ 2003]{don03}).  Although we cannot 
be certain whether or not the 
magnetic field structures extrapolated from surface magnetograms of young main sequence stars do 
represent the magnetically confined coronae of T Tauri stars, they do satisfy the currently 
available observational constraints.  In future it will be possible to use real T Tauri magnetograms 
derived from Zeeman-Doppler images obtained using the ESPaDOnS instrument at the Canada-France-Hawaii 
telescope (\cite[Donati, Jardine, Gregory, \etal\ 2007]{don07}).  However, in the meantime, the 
example field geometries used in this work 
(see Fig. \ref{coronae}) capture the essential features of T Tauri coronae.  They reproduce X-ray 
emission measures (EMs) and coronal densities which are typical of T Tauri stars 
(\cite[Jardine, Cameron, Donati, \etal\ 2006]{jar06}).  The 
surface field structures are complex, consistent with polarisation measurements 
(\cite[Valenti \& Johns-Krull 2004]{val04}) and X-ray emitting plasma is confined within unevenly 
distributed magnetic structures close to the stellar surface, giving rise to significant rotational 
modulation of X-ray emission (\cite[Gregory, Jardine, Cameron, \etal\ 2006b]{gre06b}).  


\subsection{The coronal field}
For a given surface magnetogram we calculate the extent of the closed corona for a specified
set of stellar parameters.  We extrapolate from surface magnetograms by assuming that the magnetic 
field $\boldsymbol{B}$ is potential such that $\bnabla \times \boldsymbol{B} = 0$.  This process 
is described in detail by \cite[Jardine \etal\ (2006)]{jar06}, 
\cite[Gregory, Jardine, Simpson, \etal\ (2006a)]{gre06a} and  
\cite[Gregory \etal\ (2006b)]{gre06b}.  We assume that the corona is isothermal and that 
plasma along field line loops is in hydrostatic equilibrium.  
The pressure is calculated along the path of field line loops and is set to zero for open field 
lines and for field lines where, at some point along the loop, the gas pressure exceeds the 
magnetic pressure.  The pressure along a field line scales with the pressure at its 
foot point, and we assume that this scales with the magnetic pressure.      
This technique has been used successfully to calculate mean coronal densities and X-ray EMs for 
the Sun and other main sequence stars (\cite[Jardine, Wood, Cameron, \etal\ 2002]{jar02}) as well as 
T Tauri stars (\cite[Jardine \etal\ 2006]{jar06}).  The AB Dor-like coronal field has an 
X-ray EM\footnote{The X-ray EM is given by ${\rm EM} =\int n^2dV$ where 
$n$ and $V$ are the coronal density and volume.  The EM-weighted density is 
$\bar{n}=\int n^3dV/\int n^2dV$.} of $\log{{\rm EM}}=53.73\,{\rm cm}^{-3}$ (without considering accretion) 
and a mean EM-weighted coronal density of $\log{\bar{n}}=10.57\,{\rm cm}^{-3}$, consistent with 
estimates from the modelling of individual flares 
(\cite[Favata, Flaccomio, Reale, \etal\ 2005]{fav05}).  The LQ Hya-like field has a more 
extended corona and consequently a lower coronal density and EM, 
$\log{{\rm EM}}=52.61\,{\rm cm}^{-3}, \log{\bar{n}}=9.79\,{\rm cm}^{-3}$.

\subsection{The accreting field}
We assume that the structure of the magnetic field remains undistorted by the in-falling 
material and that the magnetosphere rotates as a solid body.  The accreting field geometries 
shown in Fig. \ref{coronae} are therefore only snap-shots in time, and in reality will evolve due 
to the interaction with the disk.  The question of where the disk is truncated 
remains a major problem for accretion models.   It is still unknown if the disk is truncated in the 
vicinity of the corotation radius, the assumption of traditional accretion models (e.g. 
\cite[K{\"o}nigl 1991]{kon91}), or whether it extends closer to the stellar surface 
(e.g. \cite[Matt \& Pudritz 2005]{mat05}).  
In this work we assume that accretion occurs over a range of 
radii within the corotation radius.  This is equivalent to the 
approach taken previously by e.g. \cite{muz01} who have demonstrated that such an 
assumption reproduces observed spectral line profiles and variability.  The accretion filling factors 
are of order $1\%$, consistent with observationally inferred values (e.g. 
\cite[Valenti \& Johns-Krull 2004]{val04}).    

We assume that material is supplied by the disk and accretes onto the star at a constant rate.
For a dipolar magnetic field accretion flows impact the stellar surface in two rings in opposite 
hemispheres centred on the poles.  In this case, half of the mass supplied by the disk accretes into each
hemisphere.  For complex magnetic fields accretion occurs into discrete hot spots 
distributed in latitude and longitude (\cite[Gregory \etal\ 2006a]{gre06a}).  It is therefore not clear how 
much of the available mass from the disk accretes into each hot spot.  We use a spherical grid 
and assume that each grid cell within the disk which is accreting supplies a mass accretion 
rate that is proportional to its surface area.  If an accreting grid cell has a 
surface area that is $2\%$ of the total area of all accreting grid cells, then this grid cell 
is assumed to carry $2\%$ of the total mass that is supplied by the disk.  Therefore, as an 
example, if grid cells which constitute half of the total area of all accreting cells in the 
disk carry material into a single hot spot, then half of the mass accretion rate is carried from 
the disk to this hot spot.  In this way the accretion rate into each hot spot is different and 
depends on the structure of the magnetic field connecting the star to the disk.  


\subsection{Accretion flow model}
\begin{figure}
\centerline{
\scalebox{0.38}{%
\includegraphics{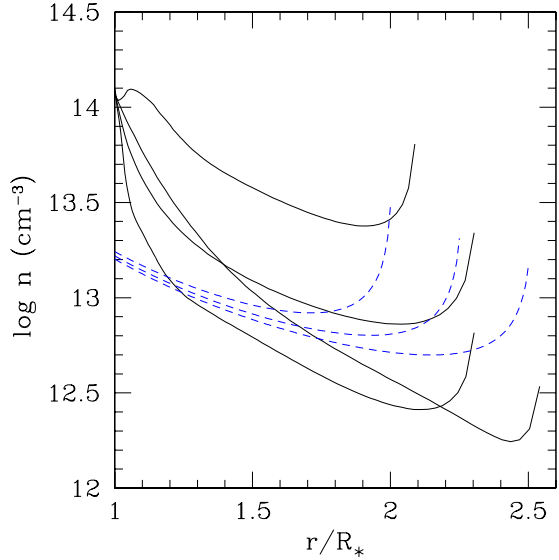}%
}
}
\caption{Some example density profiles (solid lines) for accretion along a small selection of the
        complex field lines shown in Fig. \ref{coronae} (first column, third row) assuming a mass accretion
        rate of $10^{-7}\,{\rm M}_{\odot}\,{\rm yr}^{-1}$.  Also shown for comparison are the 
        density profiles for accretion along dipolar field lines (dashed lines). $r$ is the spherical 
        radius.  Reproduced from \cite[Gregory \etal\ (2007)]{gre07}.}
    \label{flows}
\end{figure}

We consider a star of mass $0.5\,{\rm M}_{\odot}$, radius $2\,{\rm R}_{\odot}$,
rotation period $6\,{\rm d}$, a coronal temperature of $20\,{\rm MK}$ and assume
that the disk supplies a mass accretion rate of $10^{-7}\,{\rm M}_{\odot}\,{\rm yr}^{-1}$.
In order to model the propagation of coronal X-rays through the magnetosphere 
we first need to determine the density of gas within accretion columns.  
\cite[Gregory, Wood \& Jardine (2007)]{gre07} develop a steady state accretion flow model where material 
accretes from a range of radii within corotation, free-falling 
along the field lines under gravity.  The resulting density profiles do not depend on the 
absolute field strength, but 
instead on how the field strength varies with height above the star.  The density profiles are 
typically steeper than those derived for accretion flows along dipolar field lines since the 
strength of a higher order field drops faster with height above the star.
Fig. \ref{flows} shows the variation of the number density along the paths of a selection of 
accreting field lines, with 
those obtained for dipolar field lines shown for comparison (\cite[Gregory \etal\ 2007]{gre07}).  
For our assumed accretion rate of $10^{-7}\,{\rm M}_{\odot}\,{\rm yr}^{-1}$ the flow densities range from 
$\log {n} \approx 12-14\,{\rm cm}^{-3}$, whilst for a lower accretion rate of 
$10^{-8}\,{\rm M}_{\odot}\,{\rm yr}^{-1}$ the range is $\log {n} \approx 11-13\,{\rm cm}^{-3}$.  

\section{Simulated X-Ray Variability}
We model the propagation of coronal X-rays through the magnetosphere
by considering absorption by the dense gas within accretion columns.  For the 
radiation transfer we use Monte Carlo techniques and discretise the emissivity 
and density onto a spherical polar grid  (e.g. \cite[Jardine \etal\ 2002]{jar02};
\cite[Whitney, Wood, Bjorkman, \etal\ 2003]{whi03}).  The stellar inclination is set to $i=60\,^{\circ}$ 
and we assume the X-ray emission from the $20\,{\rm MK}$ corona is optically thin, but that
the X-rays may be subsequently absorbed in the cool and hence
optically thick accretion columns.  For the X-ray absorptive opacity we
adopt a value of $\sigma = 10^{-22}\,{\rm cm}^2\,{\rm H}^{-1}$,
typical of neutral gas at temperatures below $10^4\,{\rm K}$ at X-ray energies
of a few keV (e.g. \cite[Krolik \& Kallman 1984]{kro84}).  At these energies the
opacity of hot gas (above $10^7\,{\rm K}$) is several orders of magnitude
lower (e.g. \cite[Krolik \& Kallman 1984]{kro84}, their Fig. 1) justifying our
assumption that the coronal X-ray emission is optically thin.

In the Monte Carlo X-ray radiation transfer simulations we assume the
scattering opacity is negligible, so our results in Fig. \ref{coronae} show the
effects of attenuation of the coronal emission by the accretion
columns.  The second row in Fig. \ref{coronae} shows the X-ray images in the 
absence of attenuation (i.e. X-ray opacity in the accretion columns is set to zero) 
whilst the fourth row shows the same X-ray emission models, but with our adopted value
for the soft X-ray opacity in the accretion columns.
The observed X-ray EM is reduced by a factor of 1.4 (2.0) for the AB Dor-like (LQ Hya-like) field 
when accretion flows are considered, where the reduction factor is the average for an entire rotation 
cycle.  For the AB Dor-like field there are large accretion curtains which cross the observers 
line-of-sight to the star as it rotates (see Fig. \ref{coronae}).  For the LQ Hya-like field 
accretion is predominantly along field lines which carry material into low latitude hot spots, 
however, one of the brightest X-ray emitting regions is obscured by an accretion column which 
attenuates the coronal X-rays and produces a large reduction in the observed X-ray emission.  
This immediately suggests that the geometry of the accreting field is a contributory factor 
in causing the large scatter seen in the X-ray luminosities of accreting stars.  


\section{Summary}
We have demonstrated that the suppression of X-ray emission in accreting stars 
apparent from CCD observations can, at least in part, be explained by the 
attenuation of coronal X-rays by the dense material in accretion columns.   This suggests that 
both accreting and non-accreting
stars have the same intrinsic X-ray luminosity, with accreting T Tauri stars 
being observed to be less luminous due 
to the effects of absorbing gas in accretion columns.  The 
reduction in the observed X-ray emission depends on the structure of the 
accreting field.  For stars where accretion columns rotate across the 
line-of-sight, X-rays from the underlying corona are strongly absorbed by 
the accreting gas which reduces the observed X-ray emission.  A preliminary calculation 
indicates that the column densities from our simulations are large enough 
that the softer (cooler) coronal spectral components may be substantially, if not completely, absorbed
by the accreting gas.  The effect is greater the larger the accretion rate.  Indeed \cite[G{\"u}del \etal\ (2007b)]{gue07b}
have recently found that stars in XEST which have the largest accretion rates (and are driving jets) show a highly
absorbed coronal spectral component, which is attributed to attenuation by accreting gas. 
This however does not rule out the fact that other mechanisms 
may also be responsible for reducing the X-ray emission in accreting 
stars.  \cite[Jardine \etal\ (2006)]{jar06} 
have demonstrated that some stars (typically those of lower mass) have their outer coronae stripped 
away via the interaction with a disk.  This also reduces the observed X-ray 
emission and this effect, combined with the radiative transfer calculations 
presented here, is likely to lead to a larger reduction in the observed 
X-ray emission.  This would reduce the number of field lines which could be 
filled with coronal gas, such as is also suggested by \cite[Preibisch \etal\ (2005)]{pre05} and 
\cite[Telleschi \etal\ (2007a)]{tel07a}, with the observed X-ray emission being further reduced due to 
obscuration by the accreting gas.


\begin{discussion}

\discuss{Ardila}{Is absorption not corrected for when calculating X-ray luminosities?}

\discuss{Gregory}{The attenuation of X-rays is by gas in accretion columns, not by dust.  Thus the absorption is larger 
than would be calculated from say the optical extinction.  Indeed there is already evidence
that for some stars the gas-to-dust ratio is larger than what is normally assumed, leading to a heavily absorbed coronal spectral
component (for example, some stars in the XEST project).  Although with the caveat that such stars have some of the largest 
inferred accretion rates, and therefore we may expect more X-ray attenuation by denser accretion columns.  We are currently working
on this with Ettore Flaccomio.}

\discuss{Ardila}{So do you think the difference in the observed X-ray luminosities would disappear if 
X-ray attenuation is accounted for properly?}

\discuss{Gregory}{Yes, I believe so.}

\discuss{Kastner}{I think the reason that you're getting disbelieving comments is that I'm unsure why absorption should modify 
the X-ray luminosity since it should be accounted for already.}

\discuss{Flaccomio}{The derived column densities in the simulations can be high, which suggests that in accreting stars there may
be a cool component that is completely or substantially absorbed and so is not detected in the spectrum.}

\discuss{Johns-Krull}{Would accretion columns rotating across the line-of-sight produce detectable sharp drops in 
X-ray light curves?}

\discuss{Gregory}{Modulation 
due to bright regions entering eclipse produces a much smoother variation with rotation phase than that due to 
accretion columns rotating across the line-of-sight. However, the problem with testing that is you require X-ray observations that span at least a couple of stellar 
rotation periods, which are difficult to get observing time for.}

\discuss{Johns-Krull}{But can't you use the COUP dataset for that?}

\discuss{Gregory}{Yes, although in the COUP paper on rotational modulation of X-ray emission they looked to see if the modulation
occurred preferentially in accreting or non-accreting stars.  However, they could not say anything conclusive as most stars are too heavily
absorbed to have been studied spectroscopically from which their accretion status could have been determined.}

\discuss{Bouvier}{AA Tau may be an exception to your model.  We find that the accretion hot spot, the accretion column and the disk warp
exist at the same rotation phase, but if you look at the poster by Grosso, during an eclipse by the disk warp we saw an increase in X-ray
emission.}

\discuss{Gregory}{Perhaps AA Tau is an exception, or perhaps the increase in X-ray emission is accretion related rather than coronal 
in origin.  I need to think about AA Tau in more detail.}

\discuss{Matt}{If half of the X-ray luminosity goes into heating the accretion columns then you may expect a correlation between
X-ray luminosity and the flux in lines which form in the accretion columns.  Has anyone looked for this?}

\discuss{Gregory}{I'm not aware that anyone has looked for that.}

\discuss{Stelzer}{The reduction factor was the average for a complete rotation cycle, so it can 
be higher over a smaller rotation phase?}

\discuss{Gregory}{Yes, it can be higher, or less, depending on the field geometry and the portion of 
the rotation cycle observed.}

\end{discussion}


\end{document}